\documentclass[10pt,a4paper,twocolumn]{article}

\usepackage[T1]{fontenc}
\usepackage[utf8]{inputenc}
\usepackage{newtxtext,newtxmath}
\usepackage[margin=18mm,columnsep=7mm]{geometry}
\usepackage{setspace}
\usepackage{booktabs}
\usepackage{array}
\usepackage{multirow}
\usepackage{threeparttable}
\usepackage{amsmath}
\usepackage{graphicx}
\usepackage{float}
\usepackage{flafter}
\usepackage{tikz}
\usepackage{pgfplots}
\pgfplotsset{compat=1.18}
\usetikzlibrary{arrows.meta,positioning,fit,calc,decorations.pathreplacing}
\usepackage[numbers,sort&compress]{natbib}
\usepackage[hidelinks]{hyperref}
\usepackage{xcolor}

\newcommand{\ttitle}{Nesting or resolution? Facility proxies for variance
estimation versus confounding adjustment in claims data without facility
identifiers}

\usepackage{titlesec}
\titleformat*{\section}{\large\bfseries}
\titleformat*{\subsection}{\normalsize\bfseries}
\titleformat*{\subsubsection}{\normalsize\itshape}
\begin{document}

\twocolumn[{
\begin{center}
{\Large\bfseries \ttitle}
\end{center}
\vspace{2mm}
{\centering
Yuta Nonomiya\textsuperscript{1,2,\dag,*},
Tomoka Matsuura\textsuperscript{3,\dag},
Ryota Kawai\textsuperscript{4},
Andreas Dengel\textsuperscript{2,5},
Koichi Kise\textsuperscript{1},
Wakaba Fukushima\textsuperscript{3},
Ayumi Shintani\textsuperscript{4}

\vspace{1em}
{\small
\noindent\textsuperscript{1}~Graduate School of Informatics, Osaka Metropolitan University\\
\textsuperscript{2}~German Research Center for Artificial Intelligence (DFKI)\\
\textsuperscript{3}~Department of Public Health, Graduate School of Medicine, Osaka Metropolitan University\\
\textsuperscript{4}~Department of Medical Statistics, Graduate School of Medicine, Osaka Metropolitan University\\
\textsuperscript{5}~Department of Computer Science, RPTU Kaiserslautern-Landau\\
\textsuperscript{\dag}~These authors contributed equally to this work.\\
\textsuperscript{*}~Corresponding author.
}\par}
\vspace{3mm}
{\small \noindent\textbf{Corresponding author.} Yuta Nonomiya (nonomiya.yuta@omu.ac.jp)\par}
\vspace{4mm}
\section*{Abstract}

\noindent\textbf{Background.}
In claims databases, facility identifiers are often removed through
de-identification, leaving only coarse attributes such as bed size. Yet
facility information is needed for variance estimation, confounding
adjustment and assessment of treatment availability. We examined whether
accreditation-gated billing items can build strata standing for hospitals,
and how to construct a facility proxy for each use.

\noindent\textbf{Methods.}
The study population comprised 1,730,736 adults undergoing surgery under
general anaesthesia in the Medical Data Vision database, Japan (February
2019--July 2023). Marker-pattern strata were built from 66
accreditation-gated receipt codes, assuming that these codes are observed
with imperfect sensitivity and
near-perfect specificity. Their reflection of hospital characteristics was
measured by the variance explained ($R^2$) in ten variables. Against a
true facility identifier in an insurer-based database (DeSC Healthcare) and
in simulation with known hospital membership, marker-based groupings and
supplied-attribute cells were compared on the type I error rate for a null exposure, the confounding
removed when accreditation for a withheld code served as the confounder,
and the proportion of patients in groups mixing accredited and
non-accredited hospitals.

\noindent\textbf{Results.}
The $R^2$ of the strata fell as variables became more patient- than
hospital-determined, from 0.921 for
accreditation for generic drug use to 0.036 for patient sex. For variance
estimation, the supplied-attribute cells, which can merge but never split
hospitals, kept the type I error rate at 4.8\% with a wild cluster bootstrap,
against 5.4\% for the true identifier and 39\% when clustering was ignored.
Every finer marker-based grouping, which can split hospitals, did worse. For
adjustment, the marker pattern removed 51\% of the confounding, against 8\%
for the cells. For treatment availability, 49\% of
patients were in mixed groups under the marker pattern, against 98\% under
the cells.

\noindent\textbf{Conclusions.}
No single facility proxy serves every purpose: nesting governs variance
estimation, and resolution governs adjustment for accreditation-linked
facility characteristics and assessment of treatment availability. Where an
analysis requires adjustment for, or clustering on, the treating hospital but
no identifier is supplied, the supplied-attribute cells can carry the standard
errors with a wild cluster bootstrap, and marker-pattern strata are worth considering for the other two uses.

\vspace{1em}
\noindent\textbf{Keywords.} facility proxy; administrative claims data;
hospital accreditation; cluster-robust standard errors; facility-level
confounding; data quality
\vspace{6mm}
}]

\section{Background}

\label{pk:0}

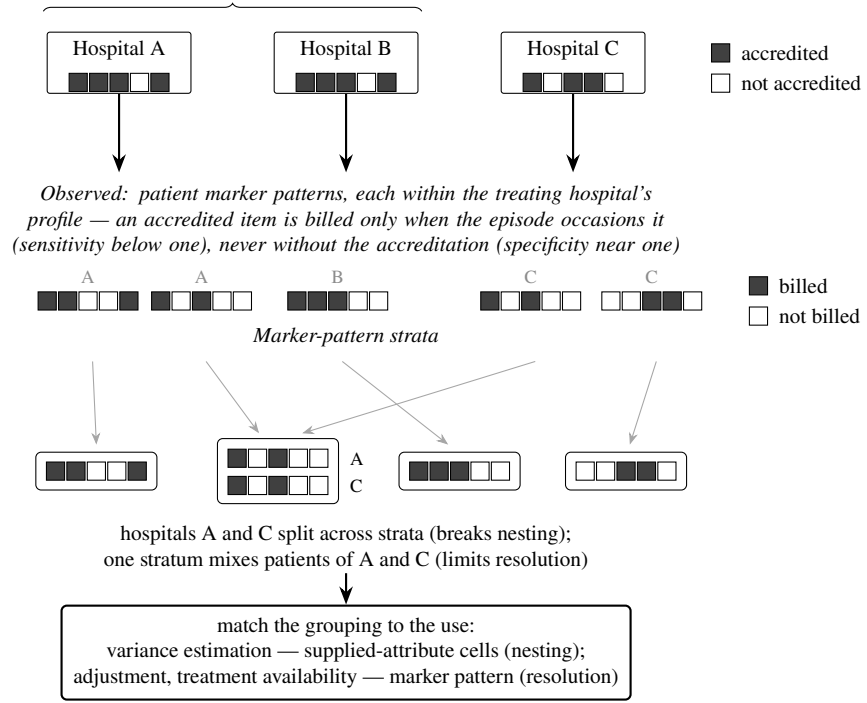
\begin{figure*}[t]
\centering
\renewcommand{\baselinestretch}{1}\selectfont
\begin{tikzpicture}[
  x=1cm, y=1cm,
  font=\fontsize{8}{9.6}\selectfont,
  hosp/.style={draw, rounded corners=1pt, inner sep=3pt, align=center, minimum width=1.9cm},
  stratum/.style={draw, rounded corners=2pt, inner sep=4pt},
  slot/.style={draw, minimum size=2.4mm, inner sep=0pt, outer sep=0pt},
  bandlab/.style={align=left, font=\fontsize{8}{9.6}\selectfont\itshape},
  flow/.style={-{Stealth[length=2.4mm]}, thick},
  assign/.style={-{Stealth[length=1.6mm]}, gray!70, thin},
  note/.style={align=left, font=\fontsize{8}{9.6}\selectfont},
]

\newcommand{\strip}[3]{
  \foreach \b [count=\i from 0] in {#3} {
    \ifnum\b=1
      \node[slot, fill=black!75] at ({#1 + \i*0.27}, {#2}) {};
    \else
      \node[slot, fill=white] at ({#1 + \i*0.27}, {#2}) {};
    \fi
  }
}

\node[bandlab, anchor=center, align=center] at (4.2, 1.75) {Unobserved: accreditation profiles of hospitals,
  filed with regional bureaus};

\draw[decorate, decoration={brace, amplitude=3pt}] (0.2, 0.68) -- (5.2, 0.68);
\node[note, anchor=south] at (2.7, 0.86)
  {identical filings: indistinguishable in principle};

\node[hosp] (hA) at (1.2, 0) {Hospital A\\[2pt]\phantom{x}};
\strip{0.66}{-0.26}{1,1,1,0,1}
\node[hosp] (hB) at (4.2, 0) {Hospital B\\[2pt]\phantom{x}};
\strip{3.66}{-0.26}{1,1,1,0,1}
\node[hosp] (hC) at (7.2, 0) {Hospital C\\[2pt]\phantom{x}};
\strip{6.66}{-0.26}{1,0,1,1,0}

\node[slot, fill=black!75] at (9.15, 0.13) {};
\node[note, anchor=west] at (9.3, 0.13) {accredited};
\node[slot, fill=white] at (9.15, -0.27) {};
\node[note, anchor=west] at (9.3, -0.27) {not accredited};

\draw[flow] (hA.south) -- (1.2, -1.45);
\draw[flow] (hB.south) -- (4.2, -1.45);
\draw[flow] (hC.south) -- (7.2, -1.45);

\node[bandlab, anchor=north, text width=10cm, align=center,
  execute at begin node={\hyphenpenalty=10000\exhyphenpenalty=10000}] at (4.2, -1.5)
  {Observed: patient marker patterns, each within the treating hospital's profile ---
   an accredited item is billed only when the episode occasions it
   \mbox{(sensitivity below one)}, never without the accreditation
   \mbox{(specificity near one)}};

\strip{0.25}{-3.15}{1,1,0,0,1}
\strip{1.75}{-3.15}{1,0,1,0,0}
\strip{3.55}{-3.15}{1,1,1,0,0}
\strip{6.10}{-3.15}{1,0,1,0,0}
\strip{7.70}{-3.15}{0,0,1,1,0}
\foreach \x/\h in {0.79/A, 2.29/A, 4.09/B, 6.64/C, 8.24/C}
  \node[font=\fontsize{7}{8.4}\selectfont, gray!90] at (\x, -2.83) {\h};

\node[slot, fill=black!75] at (9.65, -2.95) {};
\node[note, anchor=west] at (9.8, -2.95) {billed};
\node[slot, fill=white] at (9.65, -3.35) {};
\node[note, anchor=west] at (9.8, -3.35) {not billed};

\node[bandlab, anchor=center] at (4.2, -3.62) {Marker-pattern strata};

\node[stratum, minimum width=1.6cm, minimum height=0.5cm] (s1) at (0.9, -5.4) {};
\strip{0.36}{-5.4}{1,1,0,0,1}
\node[stratum, minimum width=1.6cm, minimum height=0.86cm] (s2) at (3.3, -5.4) {};
\strip{2.76}{-5.23}{1,0,1,0,0}
\strip{2.76}{-5.59}{1,0,1,0,0}
\node[font=\fontsize{7}{8.4}\selectfont, anchor=west] at (4.13, -5.23) {A};
\node[font=\fontsize{7}{8.4}\selectfont, anchor=west] at (4.13, -5.59) {C};
\node[stratum, minimum width=1.6cm, minimum height=0.5cm] (s3) at (5.7, -5.4) {};
\strip{5.16}{-5.4}{1,1,1,0,0}
\node[stratum, minimum width=1.6cm, minimum height=0.5cm] (s4) at (7.9, -5.4) {};
\strip{7.36}{-5.4}{0,0,1,1,0}

\draw[assign] (0.85, -3.95) -- (0.9, -5.05);
\draw[assign] (2.35, -3.95) -- (3.05, -4.9);
\draw[assign] (6.70, -3.95) -- (3.62, -4.9);
\draw[assign] (4.15, -3.95) -- (5.55, -5.05);
\draw[assign] (8.30, -3.95) -- (7.95, -5.05);

\node[note, anchor=north, align=center] at (4.2, -6.0)
  {hospitals A and C split across strata (breaks nesting);\\
   one stratum mixes patients of A and C (limits resolution)};

\node[draw, thick, rounded corners=2pt, inner sep=4.5pt, align=center] (inf) at (4.2, -7.8)
  {match the grouping to the use:\\
   variance estimation --- supplied-attribute cells (nesting);\\
   adjustment, treatment availability --- marker pattern (resolution)};
\draw[flow] (4.2, -6.75) -- (inf.north);

\path (-2.75, 0) -- (-2.75, 0);

\end{tikzpicture}
\caption{Building marker-pattern strata from markers and the use-specific
choice of grouping.}
\label{fig:concept}
\end{figure*}

In claims databases, facility identifiers are often removed through
de-identification. In Japan, one of the largest commercially available
hospital-based claims databases, for example, supplies no facility
identifier~\citep{iwagami2024}. The
loss of hospital identifiability matters because a facility variable,
though not the target of inference, serves at least three uses. It is
needed for variance estimation, because outcomes of inpatient care are
correlated within hospitals~\citep{localio2001,merlo2006}; for adjustment,
because hospital practice can shape the exposure and so make the hospital a
confounder~\citep{localio2001}; and for assessment of treatment
availability, because some hospitals cannot provide the exposure at all.

\label{pk:1}

When the facility variable is unavailable, each of these uses is
affected. Ignoring the clustering of outcomes
within hospitals misstates standard errors. When the exposure differs
mainly between hospitals, it understates them, so the type I error rate exceeds
its nominal level~\citep{ntani2021}. Adjustment for patient
characteristics without facility characteristics leaves facility-level
confounding in the estimated effect~\citep{localio2001}. Comparing exposed with unexposed
patients without regard to the accreditation of each patient's hospital
mixes into the unexposed group patients who could not have been exposed.

\label{pk:2}

Nevertheless, studies in such databases often set the facility aside and
report its absence only as a limitation~\citep{takano2025,iwagami2024},
thereby incurring the consequences above. The facility fields that remain
after de-identification, such as bed size, take only a handful of values,
each covering many hospitals, and studies that stratify or match on the
bed-size field acknowledge that hospitals of similar size still
differ~\citep{noda2025,nakamoto2026,bito2026}. In the United States,
registry--claims linkages have rebuilt hospital correspondence from
indirect identifiers~\citep{hammill2009}, but this requires a second data
source. Without such a source, the facility variable has to be built from
the claims themselves, as a facility proxy: a grouping of patients intended
to stand for their hospitals. Such a proxy faces a trade-off.
The supplied fields merge dozens of hospitals in each level, whereas a
finer grouping risks splitting hospitals, and misspecified clusters can
yield worse inference than none~\citep{desai2013}. How a facility proxy
should be constructed for each use in claims data without facility
identifiers has not been examined.

\label{pk:3}

This study turns to facility accreditation, as recorded indirectly in
billing items, as a new source of a facility proxy in claims data. In
Japan, hospitals
are accredited for individual services under facility-level standards on
staffing, equipment and function, and the accreditations held by each
hospital are listed in public registers~\citep{kouseikyoku2026}.
The claims do not record accreditation itself, but items that can be
billed only by accredited hospitals appear in them. Because accreditation
attaches to the hospital and changes slowly, each record can carry part of
the treating hospital's accreditation profile.

\label{pk:4}

This paper proposes a framework for building strata from such
accreditation-gated billing items (Figure~\ref{fig:concept}) and
demonstrates it in Japanese claims databases. The aims are to examine whether the resulting strata
stand for hospitals rather than patients, to determine how a facility
proxy should be constructed for each of the three uses by evaluating
candidate groupings in another database in which the true hospital is
known, and to examine the conditions
under which the framework applies to another database.

\label{pk:5}

\section{Methods}

\label{pk:6}

This section describes the two databases, the framework, stated
independently of either database, and its implementation in this study.
The candidate groupings were evaluated in the application database, in
which the hospital is unknown, and in the validation database, in which it
is known (empirical evaluation); in synthetic hospitals drawn from public
registers, with the sensitivity and specificity of marker observation
varied (Simulation 1); and in the real hospitals and marker patterns of the
validation database (Simulation 2). An example analysis of a real research
question was conducted in the application database.

\label{pk:7}

\subsection{Data sources and study population}

\label{pk:8}

We examined the framework in two settings drawn from independent claims
databases: an application setting and a validation setting. The application
setting is a commercially available hospital-based claims database (Medical
Data Vision Co., Ltd.~\citep{laurent2022}; extract February 2019--July 2023)
supplied in the format of the DPC system (Diagnosis Procedure Combination,
Japan's per-diem bundled payment system). Although this database has no facility identifier, it supplies three facility-related
fields: a three-level bed-size category (fewer than 200, 200--499, and 500 or
more beds), a designated cancer-hospital flag, and a data-contribution window
(the hospital's first and last months of data supply).
The validation setting is an insurer-based database (DeSC Healthcare,
Inc.~\citep{yasunaga2025desc}; extract April 2014--July 2023). It supplies an
anonymised facility identifier, which distinguishes facilities without
revealing the institutions they correspond to. It also records facility
attributes, including region, bed-size category and participation in the
DPC system.

\label{pk:9}

The application cohort, drawn from the Medical Data Vision database,
comprised 1,730,736 adults aged 18 or older undergoing surgery under general
anaesthesia with at least six months of prior records~\citep{nonomiya2025}
(selection codes in Additional file 1: Table S1). The date of the first
operation was defined as the index date.
The extract from the DeSC Healthcare database comprised 291,479 patients
billed, at any time during the extract period, for any of 35 codes specific
to designated intractable diseases~\citep{matsuura2024}. An
admission was identified by patient, facility and admission date, so that a
stay spanning months contributed one admission. The date of the first
admission from selection onward was defined as the index date, and the
facility submitting its inpatient claim as the index facility. After
patients whose index admission fell in the first month of the extract were
excluded, 132,767 patients remained as the validation cohort. Patients of
DPC facilities (96,270 patients, 1,741 facilities) were then excluded,
because these facilities are reimbursed for ward-wide supplements through
an institution-level coefficient, and the supplements are therefore not
itemised in the claims submitted to insurers. Patients of facilities with
unclassified payment status (249 patients, 26 facilities) were also excluded.
The remaining patients were all treated at facilities billing item by item
(fee-for-service); those with a non-empty marker pattern (29,841 patients,
2,258 facilities) were included in the analysis population (Additional
file 1: Figure~S1).

\label{pk:10}

\subsection{General framework}
Our proposed framework is based on the billing records meeting the following
four conditions. First, some billable items can be billed only by hospitals
accredited for them under a facility-level standard, so they can serve as
\emph{markers} of the hospital. Second, whether an accredited hospital bills
a marker for a patient depends mainly on the hospital rather than on the
patient's condition. Third, observation of the markers errs in one direction:
a marker rarely appears in the records of a hospital not accredited for it,
while an accredited hospital can leave it unbilled. Fourth, hospitals are
\emph{separable}, differing in the sets of markers they are accredited for.
Throughout, the target is not to identify the hospitals but to build a
grouping able to act in their place.

\label{pk:11}

Writing $S_h$ for the set of markers for which hospital $h$ is accredited and
$O_i$ for the set billed for patient $i$, the third condition states that
\begin{equation}
O_i \subseteq S_{h(i)}
\label{eq:subset}
\end{equation}
holds for nearly every patient. Observation is then misclassification in one
direction, with specificity near one and sensitivity below one. Patients of
one hospital can carry different patterns, and hospitals accredited for the
same markers carry the same ones. As billable items are introduced and
abolished at each revision of the fee schedule, the revision schedule (e.g.,
every two years in Japan) should be taken into consideration when building
the strata.

\label{pk:12}

\subsection{Candidate groupings}
We selected 66 receipt codes in 12 accreditation-gated fee categories as
markers (Additional file 1: Table S2). The categories cover medical records
management, physician clerical support, patient safety, infection control,
ward pharmacy services, data submission, discharge planning support and
hospital-level designations such as clinical training and regional medical
support. We defined a patient's \emph{marker pattern} as the set of codes
billed for the patient in the index month, which is $O_i$ in
equation~\eqref{eq:subset} (Figure~\ref{fig:concept}). Three candidate
groupings were devised.

\label{pk:13}

One candidate grouping was the observed patterns themselves, crossed in the application
database with the bed-size category and the data-contribution window, whose
groups are referred to as \emph{marker-pattern strata}. Patients with an
empty pattern were grouped by bed size and contribution window alone.

\label{pk:14}

Another was estimated by latent class analysis of the binary marker
indicators~\citep{goodman1974,nylund2007,sinha2021}, to reassemble profiles
from incompletely observed patterns, in the validation database and the
simulations. With $O_{ic}=1$ if marker $c$ is in $O_i$ and 0 otherwise,
and with markers assumed independent given class, the model for $K$
classes specifies
\begin{equation}
\Pr(O_i \mid \text{class } k) = \prod_{c} \phi_{kc}^{O_{ic}}
(1-\phi_{kc})^{1-O_{ic}},
\end{equation}
where $\phi_{kc} = \Pr(O_{ic}=1 \mid \text{class } k)$. Interpreted under
equation~\eqref{eq:subset}, $\phi_{kc}$ combines the probability that
hospitals of class $k$ are accredited for marker $c$ and the probability
that an episode gives occasion to bill it. Without separating the two
components, $\phi_{kc}$ and the class proportions were estimated by the
expectation--maximisation algorithm on the frequency-weighted table of
distinct marker patterns. The $\phi_{kc}$ were initialised from the maximal
observed patterns, largest first, at 0.85 for markers in the seed pattern
and 0.05 otherwise plus $\mathcal{N}(0, 0.03^2)$ noise from a seeded
generator, with uniform class proportions, and the algorithm ran for 60
iterations with probabilities clipped to $[0.02, 0.98]$. $K$ was selected
over a geometric grid by the Bayesian information criterion,
$-2\log L + \{KC + (K-1)\}\log N$, where $C$ is the number of markers and
$N$ the number of patients; in the validation database, the fit was
repeated from ten random restarts. Each patient was assigned to the class
of highest posterior probability.

\label{pk:15}

The other grouping was a cross-classification of supplied facility
fields, including the bed-size category, the cancer-hospital flag and the
data-contribution window in the application database, referred to as the
\emph{supplied-attribute cells}.

\label{pk:16}

The \emph{nesting} of a grouping was defined as the probability that two
patients of the same hospital shared a group, and its \emph{resolution} as
the probability that two patients of a group shared a hospital. A grouping
with nesting 1 can merge hospitals but never splits one. The
supplied-attribute cells always have nesting 1. The marker-based groupings
have higher resolution but can split a hospital when observation is
incomplete. Group counts were reported with a size-based effective number of
clusters (inverse Herfindahl index)~\citep{carter2017}.

\label{pk:17}

\subsection{Empirical evaluation}
The empirical evaluation comprised three parts: the reflection of hospital
rather than patient characteristics in the application cohort; the
observation of markers in the application database; and the consistency
between markers and hospitals in the validation database.

\label{pk:18}

\subsubsection{Reflection of hospital characteristics}
In the application cohort, we regressed each of ten variables not used to
build the strata on the indicators of the marker-pattern strata and,
separately, on those of the supplied-attribute cells. The resulting $R^2$ was
compared with the $R^2$ obtained after randomly shuffling the variable's
values across patients with the strata held fixed, and its uncertainty was
summarised by a patient bootstrap. The variables, ordered by how far the
hospital determined them, were accreditation for generic drug use, recording
of a nursing-acuity assessment and recording of any laboratory value
(hospital alone); an intensive or high-care unit fee and an anaesthesia
management fee (hospital and patient); the Barthel Index at admission and
impaired consciousness at admission (patient condition); the Charlson
comorbidity index~\citep{charlson1987,quan2005} and the number of
comorbidity categories (patient comorbidity); and sex (patient alone).

\label{pk:19}

\begin{table*}[t]
\centering
\caption{Characteristics of the two cohorts.}
\label{tab:baseline}
\footnotesize
\begin{threeparttable}
\begin{tabular}{lrr}
\toprule
 & Application cohort & Validation cohort \\
 & (Medical Data Vision) & (DeSC) \\
\midrule
Facility identifier & not supplied & anonymised \\
Patients & 1,730,736 & 36,248\textsuperscript{a} \\
Female sex & 54.7\% & 55.5\% \\
Age at index admission, median (interquartile range, IQR) & 68 (52--77) & --- \\
Aged 65 or older & 57.3\% & 90.7\% \\
Charlson index, mean; proportion with 0 & 1.78; 37.9\% & --- \\
$\ge$1 marker billed in index month & 97.9\% & 82.3\% \\
Bed-size category $<$200 / 200--499 / $\ge$500 & 7.1\% / 53.9\% / 39.0\% & --- \\
Designated cancer hospital & 73.7\% & --- \\
Facilities & --- & 2,882 \\
Patients per facility, median (IQR) & --- & 3 (1--11) \\
Analysis population & full cohort & 29,841 in 2,258 facilities\textsuperscript{b} \\
\bottomrule
\multicolumn{3}{l}{\footnotesize \textsuperscript{a}\,Fee-for-service segment, to which all validation-cohort rows refer; of 291,479}\\
\multicolumn{3}{l}{\footnotesize \phantom{\textsuperscript{a}\,}patients selected on intractable-disease billing, 132,767 were admitted from selection onward}\\
\multicolumn{3}{l}{\footnotesize \phantom{\textsuperscript{a}\,}(fee-for-service 36,248, DPC 96,270, unclassified 249).}\\
\multicolumn{3}{l}{\footnotesize \textsuperscript{b}\,Fee-for-service segment with a non-empty marker pattern.}\\
\end{tabular}
\begin{tablenotes}[flushleft]\footnotesize\setlength{\labelsep}{0pt}
\item[] Note: dashes mark entries not
assessed.
\end{tablenotes}
\end{threeparttable}
\end{table*}

\begin{table*}[t]
\centering
\caption{$R^2$ of ten variables on the grouping indicators, by how far
the hospital determines them.}
\label{tab:gradient}
\small
\setlength{\tabcolsep}{5pt}
\begin{threeparttable}
\begin{tabular}{llcc}
\toprule
Determined by & Variable & Supplied-attribute & Marker-pattern \\
 & & cells & strata \\
\midrule
\multirow{3}{*}{Hospital alone}
 & Accreditation for generic drug use & 0.046 & 0.921 \\
 & Recording of a nursing-acuity assessment & 0.017 & 0.684 \\
 & Recording of any laboratory value & 0.113 & 0.648 \\
\midrule
\multirow{2}{*}{Hospital and patient}
 & Intensive or high-care unit fee & 0.014 & 0.486 \\
 & Anaesthesia management fee & 0.071 & 0.314 \\
\midrule
\multirow{2}{*}{Patient condition}
 & Barthel Index at admission & 0.016 & 0.191 \\
 & Impaired consciousness at admission & 0.006 & 0.131 \\
\midrule
\multirow{2}{*}{Patient comorbidity}
 & Charlson comorbidity index & 0.021 & 0.116 \\
 & Number of comorbidity categories & 0.013 & 0.115 \\
\midrule
Patient alone & Sex & 0.003 & 0.036 \\
\bottomrule
\end{tabular}
\begin{tablenotes}[flushleft]\footnotesize\setlength{\labelsep}{0pt}
\item[] Note: operational definitions ---
generic-drug accreditation, any fee-category A243 code in the index month;
nursing-acuity, Barthel and consciousness items, discharge-abstract fields;
any laboratory value, at least one row in the laboratory-results file;
intensive or high-care unit fee, any A3xx code; anaesthesia management fee,
L009 or L010; Charlson index, Quan's ICD-10 algorithm.
Patient-bootstrap 95\% intervals span at most $\pm$0.007 and are
conditional on the fixed groupings. Permutation nulls were
0.002 for every variable. Barthel and consciousness items were recorded for
94.6\% and 99.0\% of the cohort; their values are computed on those
patients.
\end{tablenotes}
\end{threeparttable}
\end{table*}

\subsubsection{Marker observation in the application database}
As accreditation is assumed to be stable from month to month, a patient's marker pattern
should be contained in one of the maximal patterns (contained in no other)
observed in the database in the previous month. We computed the proportion
of patients whose pattern was so contained for each pair of adjacent months.
As an estimate of sensitivity, we computed within each calendar month the
ratio of the mean number of markers in a patient's pattern to the mean size
of the smallest maximal pattern containing it, and took the median over
months.

\label{pk:20}

\subsubsection{Consistency between markers and hospitals}
In the analysis population, the agreement between the marker pattern and
the true facility identifier and the sensitivity and specificity of marker
observation were evaluated. For agreement, the adjusted Rand
index~\citep{hubert1985} was computed within each of the five fee-schedule
periods and over the full extract. For sensitivity, the markers of the
66-code list billed for any patient of a hospital were pooled into its
facility union within each fee-schedule period and over the full extract.
The total number of markers in the patients' patterns was divided by the
total size of their facility unions, and the 95\% confidence interval (CI)
was obtained from 500 bootstrap samples drawn at the hospital level. For
specificity, each patient's pattern
was compared with the markers billed for the other patients of the same
hospital. The proportion of patients whose pattern was contained in those
markers was computed in hospitals with at least 10 patients and in those
with at least 30.

\label{pk:21}

\subsection{Simulation studies}
Two simulation studies evaluated the candidate groupings against the true
hospitals. Simulation 1 swept sensitivity and specificity in synthetic
hospitals, and Simulation 2 kept them as observed in the validation-database
patients. Both are reported according to the ADEMP
structure (aims, data-generating mechanisms, estimands, methods and
performance measures)~\citep{morris2019}; Additional file 2 maps each
element to the text. Unless stated otherwise, standard errors were
cluster-robust with the usual small-sample correction (CR1), and tests were
based on normal critical values ($|t| > 1.96$ at the 5\% level). The numbers of simulated cohorts and
replications and the Monte Carlo standard errors are given with each
result. Random-number streams were seeded by scenario and replication.

\label{pk:22}

\subsubsection{Simulation 1: sensitivity and specificity}
\paragraph{Aims.}
To examine how imperfect observation of the markers affects variance
estimation when the groupings are rebuilt from the observed patterns, and
how this depends on the separability of the hospitals.

\label{pk:23}

\paragraph{Data-generating mechanisms.}
Each simulated cohort drew 440 hospitals, matching the roster documented at
the time of the cohort extraction (about one quarter of acute-care
hospitals in Japan~\citep{nonomiya2025}), from a catalogue of accreditation profiles
assembled from public registers downloaded in July 2026, after the
extraction window~\citep{kouseikyoku2026}, and allocated
132,000 patients to them in proportion to bed count. The outcome and the
null exposure were
\begin{equation}
y_i = u_{h(i)} + e_i, \qquad
x_i = \mathbf{1}\{\sqrt{0.5}\,v_{h(i)} + \sqrt{0.5}\,\varepsilon_i > 0\},
\end{equation}
with $u_h \sim \mathcal{N}(0,\,0.14/0.86)$ giving an outcome intraclass
correlation of 0.14, and $v_h$, $e_i$ and $\varepsilon_i$ independent
standard normal, so that half of the latent exposure variation lay between
hospitals and the exposure had no effect. Each marker in the treating
hospital's profile entered the patient's pattern independently with
probability equal to the sensitivity, which ranged from 0.40 to 1.00 with
no marker observed outside the profile. At sensitivity 0.85, specificity
was lowered by letting 1\% and 5\% of markers be billed without
accreditation. The intraclass correlation was also varied (0.05, 0.14 and
0.30) at sensitivities 0.742 and 1.00 (Additional file 1: Table S3).
Separability was computed from the registers as the mean number of
hospitals per distinct profile, nationally over all hospitals with at least
20 beds holding at least one marker and, at the scale of the application
cohort, over 20 random draws of 440 hospitals with at least 100 beds. To vary separability, the 20 inpatient
notification names that make up the simulation catalogue were thinned to 16, 12, 8, 6 and 4 before the hospitals
were drawn, with 20 random subsets at each size, at sensitivities 1.00 and
0.742 (Additional file 1: Table S4). The supplied-attribute cells crossed
each hospital's bed-size category with a data-contribution window drawn
from those observed in the application database.

\label{pk:24}

\paragraph{Estimands.}
The effect of the exposure on the outcome, whose true value was zero.

\label{pk:25}

\paragraph{Methods.}
The effect was estimated by the coefficient of the exposure in the linear
regression of the outcome on the exposure. The marker-based groupings were
rebuilt from the generated patterns in each cohort. The following choices of clustering for the standard errors were
compared~\citep{desai2013}: no clustering, the marker pattern, the latent classes, the supplied-attribute cells (with
normal critical values and with a wild cluster
bootstrap~\citep{cameron2015,mackinnon2017}), and the true hospitals.
Fixed effects for the marker-based groupings and the true hospitals were
crossed with the clustering of the standard errors (Additional file 1: Table S5).

\label{pk:26}

\paragraph{Performance measures.}
The empirical type I error rate for the null exposure at the nominal 5\% level
with its Monte Carlo standard error. Under thinned catalogues, the number of distinct profiles among the drawn
hospitals, the mean number of hospitals per profile, the share of hospitals
holding no marker and the adjusted Rand index of each grouping against hospital membership were also recorded.

\label{pk:27}

\subsubsection{Simulation 2: the three uses}
\paragraph{Aims.}
To evaluate each candidate grouping for variance estimation, adjustment and
treatment availability against the true identifier, on real hospital
membership and real marker patterns.

\label{pk:28}

\paragraph{Data-generating mechanisms.}
The 29,841 patients of the analysis population kept their true hospitals
and their observed patterns. For variance estimation, the outcome carried a
facility random intercept and the null exposure a facility-level component,
generated as in Simulation 1. For adjustment and treatment availability,
one marker code at a time was withheld, the hospital's accreditation for
that code acted as a real facility-level confounder, and a null exposure
was generated from it at fixed facility-driven strength, over six codes
(each accredited at 24--65\% of hospitals).

\label{pk:29}

\paragraph{Estimands.}
For variance estimation, the effect of the exposure on the outcome, whose
true value was zero. For adjustment, the residual confounding after adjusting
for group fixed effects (subtracting group means from exposure and
outcome), scaled from no adjustment (0) to the true identifier (1). For
treatment availability, the withheld code represented a treatment
available only at hospitals accredited for it, and the target was the
proportion of patients whose group mixed hospitals accredited for the code
with hospitals that were not.

\label{pk:30}

\paragraph{Methods.}
The marker-based groupings were built from the pattern alone and, for
adjustment and treatment availability, rebuilt without the withheld code.
For variance estimation, the effect was estimated as in Simulation 1.
The supplied-attribute cells cross-classified region and bed-size category
(<200, 200--499, $\ge$500 beds). For comparison, cells were also formed from all seven
facility fields of the validation database (facility type, university
hospital, designated cancer hospital, emergency and critical care centre,
participation in the DPC system, region and bed-size category), and
clustering was also ignored. To separate nesting from resolution in variance
estimation, \emph{random hospital groups}, which never split a hospital but
carry no facility information, were formed by assigning each hospital at
random to one of 14 groups, 20 times. The true hospitals were also split at
random: each patient was moved with probability $q$ to a second group of
the same hospital, with $q$ chosen to give a nesting of 0.95, 0.90, 0.80,
0.70 and 0.50, and each hospital was divided into 3, 5 and 10 random parts.
For the supplied-attribute cells and the random hospital groups, the type I
error rate was also computed with a wild cluster bootstrap.

\label{pk:31}

\paragraph{Performance measures.}
The empirical type I error rate at the nominal 5\% level; the fraction of
confounding removed; and the proportion of patients in mixed groups.

\label{pk:32}

\subsection{Illustrative application}
To illustrate the groupings on a real exposure, we analysed the association
between surgery on a non-working day and postoperative stay in the
application cohort, a question studied in hip
fracture~\citep{boutera2020,mori2025}. The population comprised the 39,939
patients undergoing open reduction and internal fixation of a fracture whose
index admission began within the hospital's data-contribution window. The
exposure was surgery on a non-working day, defined as an index operation on
a Saturday, Sunday or national holiday. The outcome was the log of the
postoperative stay (discharge date minus operation date plus one), with
in-hospital deaths contributing their observed stay. Three nested linear
specifications were fitted: a crude regression without covariates;
patient-level adjustment for age (restricted cubic spline with three knots
at the 10th, 50th and 90th percentiles, 48, 78 and 92 years), sex,
emergency admission and the Charlson index; and the latter with
marker-pattern strata fixed effects added. Each was fitted with clustering
ignored and with standard errors clustered on the supplied-attribute cells
(cell-clustered errors), with the CR1 small-sample factor $G/(G-1)$ for $G$
clusters. For the cell-clustered fits, intervals and $p$-values came from a
null-imposed wild cluster bootstrap-$t$ with Rademacher weights and 999
replications. Each 95\% interval was $\hat\beta \pm q\,\mathrm{SE}$, with
$q$ the 95th percentile of the bootstrap $|t|$. We also computed the $R^2$
of the exposure on the strata indicators and counted the strata containing
both exposure levels and the patients in them.

\label{pk:33}

\subsection{Software}
All analyses were run in Python 3.12 with NumPy 2.4, SciPy 1.17, pandas 3.0
and polars 1.41.

\label{pk:34}

\subsection{Use of large language models}
The authors used large language models (Claude, Anthropic) to support the
drafting and revision of text and analysis code. These tools were not used
to generate data or results. Every reported value was produced by code the
authors ran and inspected. The authors reviewed and edited the content as
needed and take full responsibility for it.

\label{pk:35}

\section{Results}

\label{pk:36}

\subsection{Cohorts and candidate groupings}
Characteristics of both cohorts are in Table~\ref{tab:baseline}. In the
application cohort, at least one marker was billed in the index month for
97.9\% of patients. The remaining 2.1\% carried the empty pattern and were
grouped by bed size and contribution window alone. Bed size and
contribution window formed 76 combinations. Adding the cancer-hospital flag
gave 94 supplied-attribute cells, whose effective number of clusters was
8.4. The marker pattern split these 76 combinations into 16,072 strata. The
median stratum held 10 patients. The largest tenth of strata held 76.9\% of
patients. In the validation database, the Bayesian information criterion
selected 60 latent classes in each of the ten restarts.

\label{pk:37}

\subsection{Reflection of hospital characteristics}
In the application cohort, for the marker-pattern strata, the $R^2$ was 0.921 for accreditation for
generic drug use, 0.684 for recording of a nursing-acuity assessment, 0.648
for recording of any laboratory value, 0.314--0.486 for the intensive or
high-care unit fee and the anaesthesia management fee, 0.131--0.191 for the
Barthel Index and impaired consciousness at admission, 0.116 for the
Charlson comorbidity index, and 0.036 for sex (Table~\ref{tab:gradient}).
For the supplied-attribute cells, it was at most 0.113.

\label{pk:38}

\subsection{Marker observation in the application database}
In the application database, over 53 pairs of adjacent months (5,365,164 patient-months), the median
monthly proportion of patients whose pattern was contained in a maximal
pattern of the previous month was 0.989 (interquartile range 0.986--0.992).
The completeness-based estimate of sensitivity was 0.741.

\label{pk:39}

\begin{figure}[tbp]
\centering
\renewcommand{\baselinestretch}{1}\selectfont
\definecolor{cBlue}{HTML}{2A78D6}
\definecolor{cOrange}{HTML}{EB6834}
\definecolor{cInk}{HTML}{0B0B0B}
\definecolor{cInk2}{HTML}{52514E}
\definecolor{cGrid}{HTML}{E6E5E1}
\begin{tikzpicture}
\begin{axis}[
  width=8.2cm, height=6.0cm,
  xmin=0.37, xmax=1.03, ymin=0, ymax=0.72,
  xtick={0.40,0.50,0.60,0.70,0.85,0.95,1.00},
  xticklabels={0.40,0.50,0.60,0.70,0.85,0.95,1.00},
  ytick={0,0.05,0.1,0.2,0.3,0.4,0.5,0.6,0.7},
  yticklabels={0,0.05,0.10,0.20,0.30,0.40,0.50,0.60,0.70},
  tick label style={font=\scriptsize, color=cInk2},
  xlabel={Sensitivity of marker observation}, ylabel={Empirical type I error rate},
  label style={font=\footnotesize, color=cInk},
  axis line style={cInk2, -}, tick style={cInk2},
  axis lines*=left, grid=major, grid style={cGrid, line width=0.4pt},
  legend style={at={(0.98,0.86)}, anchor=north east, draw=cGrid, fill=white,
    font=\scriptsize, text=cInk, row sep=0pt, inner sep=2pt, cells={anchor=west}},
  legend cell align=left,
  clip=false,
]
\addplot[cInk2!60, line width=1.2pt, domain=0.40:1.00, samples=2, forget plot] {0.666};
\node[anchor=south west, font=\scriptsize, color=cInk2] at (axis cs:0.40,0.668) {Ignoring clustering};
\addplot[cOrange, line width=1.6pt, mark=square*, mark size=2.0pt,
  mark options={draw=white, line width=0.8pt}] coordinates {
  (0.40,0.584) (0.50,0.575) (0.60,0.563) (0.70,0.518) (0.85,0.364) (0.95,0.163) (1.00,0.063)};
\addlegendentry{Marker pattern}
\addplot[cBlue, line width=1.6pt, dash pattern=on 5pt off 2.5pt, mark=*, mark size=2.2pt,
  mark options={solid, draw=white, line width=0.8pt}] coordinates {
  (0.40,0.489) (0.50,0.384) (0.60,0.289) (0.70,0.242) (0.85,0.146) (0.95,0.097) (1.00,0.063)};
\addlegendentry{Latent classes}
\addplot[cInk, line width=1.2pt, domain=0.40:1.00, samples=2, forget plot] {0.051};
\node[anchor=south west, font=\scriptsize, color=cInk] at (axis cs:0.40,0.053)
  {Supplied-attribute cells; true hospitals};
\addplot[cInk2, dotted, line width=0.8pt, forget plot] coordinates {(0.742,0) (0.742,0.72)};
\node[anchor=south, font=\footnotesize, color=cInk2, align=center] at (axis cs:0.742,0.72)
  {sensitivity measured in\\the validation database};
\end{axis}
\end{tikzpicture}
\caption{Empirical type I error rate by sensitivity of marker observation
(Simulation 1). Nominal 5\% level; 440 hospitals; 20 simulated cohorts
$\times$ 1,000 replications per value; Monte Carlo standard errors at most
0.007. The supplied-attribute cells (bed-size category $\times$
data-contribution window; effective number of clusters 6.8) were tested with
a wild cluster bootstrap; their line and that of the true hospitals lie
within 0.002 and are drawn as one. The dotted line marks the sensitivity measured in
the validation database.}
\label{fig:sim}
\end{figure}
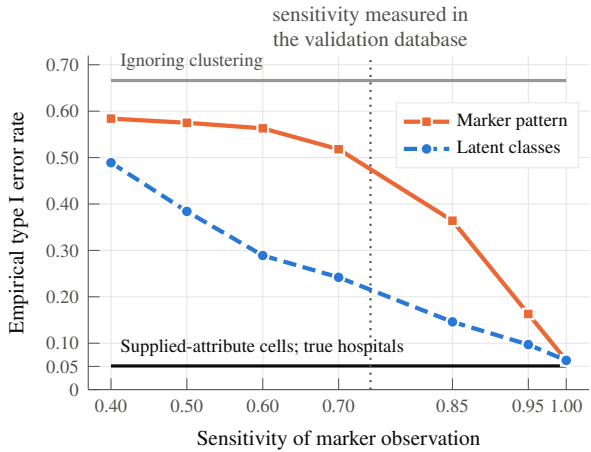

\begin{figure}[tbp]
\centering
\renewcommand{\baselinestretch}{1}\selectfont
\definecolor{cBlue}{HTML}{2A78D6}
\definecolor{cOrange}{HTML}{EB6834}
\definecolor{cInk}{HTML}{0B0B0B}
\definecolor{cInk2}{HTML}{52514E}
\definecolor{cGrid}{HTML}{E6E5E1}
\begin{tikzpicture}
\begin{axis}[
  width=8.2cm, height=6.0cm,
  xmin=-0.03, xmax=1.06, ymin=0, ymax=0.50,
  xtick={0,0.2,0.4,0.6,0.8,1.0},
  xticklabels={0,0.2,0.4,0.6,0.8,1.0},
  ytick={0,0.05,0.1,0.2,0.3,0.4},
  yticklabels={0,0.05,0.10,0.20,0.30,0.40},
  tick label style={font=\scriptsize, color=cInk2},
  xlabel={Nesting: probability that two patients\\of a hospital share a group},
  xlabel style={align=center},
  ylabel={Empirical type I error rate},
  label style={font=\footnotesize, color=cInk},
  axis line style={cInk2, -}, tick style={cInk2},
  axis lines*=left, grid=major, grid style={cGrid, line width=0.4pt},
  legend style={name=lgd, at={(0.98,0.98)}, anchor=north east, draw=cGrid, fill=white,
    font=\scriptsize, text=cInk, row sep=0pt, inner sep=2pt, cells={anchor=west}},
  legend cell align=left,
  clip=false,
]
\addplot[cBlue, line width=1.6pt, mark=*, mark size=2.0pt,
  mark options={draw=white, line width=0.8pt}] coordinates {
  (0,0.389) (0.100,0.307) (0.200,0.248) (0.334,0.189) (0.501,0.138)
  (0.701,0.094) (0.797,0.078) (0.904,0.064) (0.949,0.059) (1,0.054)};
\addlegendentry{True hospitals split at random}
\addplot[only marks, cOrange, mark=diamond, mark size=4.6pt, line width=1.0pt,
  mark options={fill=white}] coordinates {(0.425,0.168)};
\addlegendentry{Latent classes}
\addplot[only marks, cOrange, mark=square, mark size=3.2pt, line width=1.0pt,
  mark options={fill=white}] coordinates {(0.171,0.264)};
\addlegendentry{Marker pattern}
\addplot[only marks, cInk, mark=triangle, mark size=3.4pt, line width=0.8pt,
  mark options={fill=white}] coordinates {(1.03,0.091)};
\addlegendentry{Supplied-attribute cells}
\addplot[only marks, cInk, mark=o, mark size=2.4pt, line width=0.8pt,
  mark options={fill=white}] coordinates {(0.97,0.074)};
\addlegendentry{Random hospital groups}
\addplot[only marks, cInk, mark=triangle*, mark size=3.4pt, forget plot] coordinates {(1.03,0.048)};
\addplot[only marks, cInk, mark=*, mark size=2.2pt, mark options={draw=white, line width=0.6pt},
  forget plot] coordinates {(0.97,0.050)};
\addlegendimage{empty legend}
\addlegendentry{\hspace{-0.6cm}\shortstack[l]{Open, normal critical values;\\filled, wild cluster bootstrap}}
\end{axis}
\end{tikzpicture}
\caption{Empirical type I error rate against nesting (Simulation 2).
Validation database, 29,841 patients in 2,258 facilities; nominal 5\% level
with normal critical values unless stated; 20,000 replications per grouping;
Monte Carlo standard errors at most 0.0034. The line splits the true
hospitals at random with resolution held at 1, from ignoring clustering (left
end) to the true identifier (right end). Symbols mark the groupings of
Table~\ref{tab:axes} and the random hospital groups at their observed
nesting, offset horizontally at nesting 1 for legibility.}
\label{fig:nesting}
\end{figure}
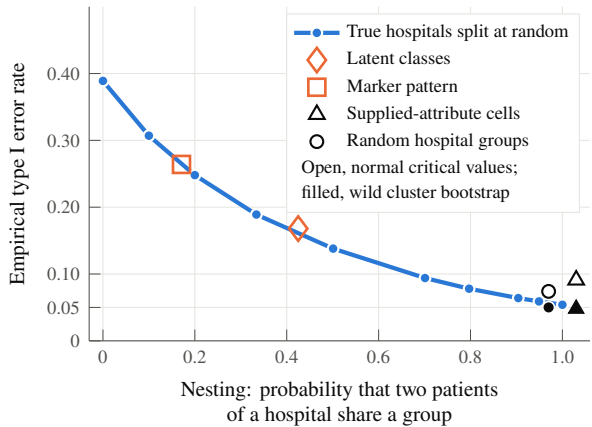

\subsection{Consistency between markers and hospitals}
In the analysis population of the validation database, sensitivity was 0.742 (95\% CI 0.722--0.761)
within a fee-schedule period, ranging from 0.649 to 0.824 across the five
periods, against 0.488 when the facility union was pooled over nine years.
The adjusted Rand index of the marker pattern against true hospital
membership was 0.108--0.345 within the five periods, against 0.098 pooled.
Within hospitals, a patient's pattern was contained in the other patients'
markers for 98.6\% of patients where the hospital had ten or more patients,
rising to 99.2\% at thirty or more.

\label{pk:40}

\subsection{Simulation 1: sensitivity and specificity}
Simulation 1 drew its hospitals from the public registers rather than from
either database. Figure~\ref{fig:sim} shows the type I error rate at the
nominal 5\% level in cohorts of 440 simulated hospitals with known
membership. Ignoring clustering
rejected in about two thirds of null datasets. Clustering on the latent
classes held 6.3--24.2\% at sensitivities 0.70--1.00 and degraded smoothly
below that range (28.9--48.9\% at sensitivities 0.40--0.60), never worse
than no correction. The marker pattern matched the latent classes only near
complete observation. The supplied-attribute cells, built without markers
and so independent of sensitivity, held 5.0\% with a wild cluster bootstrap,
against 9.8\% with normal critical values. At sensitivity 0.85, letting 1\%
of markers be billed without accreditation raised the latent-class error to
24.8\%, and 5\% took both marker-based groupings to 34.3--55.0\%.

\label{pk:41}

\begin{table*}[t]
\centering
\caption{The three uses of a facility variable, evaluated against the
true identifier.}
\label{tab:axes}
\footnotesize
\setlength{\tabcolsep}{1.4pt}
\begin{threeparttable}
\begin{tabular}{lcccc}
\toprule
 & Supplied-attribute & Latent class & Marker pattern & True identifier \\
 & cells (31 groups; & (60 groups; & (3,433 groups; & (2,258 facilities; \\
 & eff.\ 8.2) & eff.\ 19.9) & eff.\ 139.6) & eff.\ 328.3) \\
\midrule
Variance estimation & & & & \\
\enspace Type I error rate & 0.048 (0.002)\textsuperscript{b} & 0.168 (0.003) & 0.264 (0.003) & 0.054 (0.002) \\
Adjustment & & & & \\
\enspace Fraction of confounding removed & 0.084 [0.04--0.23] & 0.210 [0.13--0.29] & 0.507 [0.38--0.65] & 1.000\textsuperscript{a} \\
Availability & & & & \\
\enspace Proportion of patients in mixed groups & 0.975 [0.87--1.00] & 0.907 [0.86--0.96] & 0.494 [0.42--0.54] & 0.000 \\
\bottomrule
\end{tabular}
\begin{tablenotes}[flushleft]\footnotesize\setlength{\labelsep}{0pt}
\item[] Note: validation database, fee-for-service segment; the three uses
were evaluated jointly on the same patients. eff., effective number of
clusters (inverse Herfindahl index). Type I error rate at the nominal 5\%
level. Fraction of confounding removed: 0 = no adjustment, 1 = true
identifier. Mixed group: hospitals with and without accreditation for the
withheld code. Parentheses, Monte Carlo standard errors (20,000
replications); brackets, range over the six withheld codes, not a
confidence interval (per-code values in Additional file 1: Table S6).
Ignoring clustering: 0.389 (0.003). Cells from all seven facility fields
(109 groups; eff.\ 26.2): 0.064 (0.002), 0.147 and 0.937. Random hospital
groups (14 groups, 20 assignments; eff.\ 13.0--13.7): 0.074 with normal critical values
and 0.050 with a wild cluster bootstrap (ranges over the 20 assignments,
0.071--0.080 and 0.047--0.053).
\textsuperscript{a}\,Analytic ceiling. \textsuperscript{b}\,Wild cluster
bootstrap; 0.091 (0.002) with normal critical values.
\end{tablenotes}
\end{threeparttable}
\end{table*}

Varying the intraclass correlation from 0.05 to 0.30 left the ordering at
sensitivity 0.742 unchanged: the type I error rate was lowest for the true
hospitals, followed by the latent classes, the marker pattern and no
clustering (Additional file 1: Table S3). Fixed effects for a marker-based
grouping did not correct the unclustered test, whose error was
63.4--64.4\% with them against 65.5\% without them (Additional file 1:
Table S5).

\label{pk:42}

Nationally, the 6,993 hospitals with at least 20 beds that held at least
one marker occupied 960 distinct accreditation profiles (7.28 hospitals per
profile). Among 440 hospitals drawn from the registers, each distinct
profile was shared on average by 2.2 hospitals (199 profiles). Thinning the 20 marker names of the simulation
catalogue lowered this separability (Additional file 1: Table S4). At full
observation, with 20, 12 and 4 marker names, the hospitals per profile were
2.8, 7.4 and 63 (with 40\% of hospitals holding no marker at 4), and the
adjusted Rand index of the marker pattern against hospital membership was
0.21, 0.08 and 0.01. The type I error rate of the marker pattern rose from
5.9\% with 20 markers to 7.6\% with 12, against 4.4--5.5\% for the true
hospitals, and reached 9.6--20.2\% only at 8 or fewer markers, where the
hospitals fell into 29 or fewer profiles. At sensitivity 0.742, thinning
lowered the marker pattern's rate from 49.6\% to 32\%, while the latent
classes stayed at 21--28\% down to six markers; with four markers, too few
distinct patterns were observed in many subsets to fit the latent class
model.

\label{pk:43}

\subsection{Simulation 2: the three uses}
Simulation 2 was run on the analysis population of the validation database,
and Table~\ref{tab:axes} shows the results for the three uses. For variance
estimation, the type I error rate with normal critical values was 9.1\% for the
supplied-attribute cells, 7.4\% for the random hospital groups
(7.1--8.0\% over 20 assignments), 16.8\% for the latent classes and 26.4\%
for the marker pattern, against 5.4\% for the true identifier. With a wild
cluster bootstrap, the type I error rate was 4.8\% for the
supplied-attribute cells and 5.0\% (4.7--5.3\%) for the random hospital
groups. Figure~\ref{fig:nesting} plots the type I error rate against nesting. When
the true hospitals were split at random with resolution held at 1, the type
I error rate rose from 5.9\% at a nesting of 0.95 to 13.8\% at 0.50 and
30.7\% at 0.10, against 38.9\% when clustering was ignored. The latent classes (nesting 0.43, resolution 0.03) and the marker
pattern (nesting 0.17, resolution 0.07) lay on this curve, at 16.8\% and
26.4\%, whereas the supplied-attribute cells and the random hospital
groups, with nesting 1 and resolution 0.02--0.04, lay at 7.4--9.1\% with
normal critical values and 4.8--5.0\% with a wild cluster bootstrap.
For adjustment, the fraction of confounding removed by group fixed effects
was 51\% for the marker pattern, 21\% for the latent classes and 8\% for the
supplied-attribute cells (per withheld code: Additional file 1: Table S6).
For treatment availability, the proportion of patients in mixed groups was
98\% for the supplied-attribute cells and 49\% for the marker pattern.

\label{pk:44}

\subsection{Illustrative application}
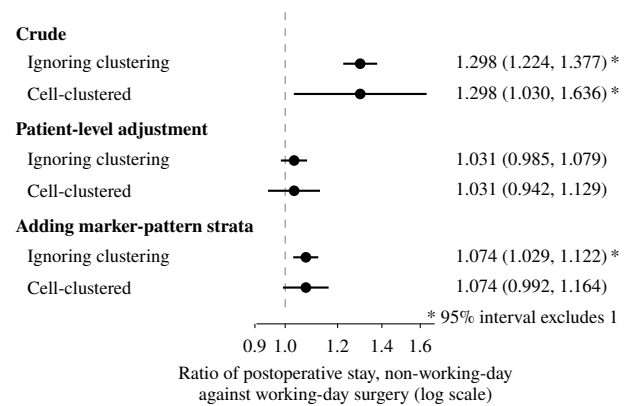
\begin{figure}[tbp]
\centering
\begin{tikzpicture}[x=3.8cm, y=0.8cm]
\draw[dashed, gray] (0, 0.35) -- (0, -4.55);
\node[anchor=west, font=\scriptsize\bfseries] at (-0.97, 0) {Crude};
\node[anchor=west, font=\scriptsize] at (-0.93, -0.5) {Ignoring clustering};
\draw[line width=0.8pt] (0.202, -0.5) -- (0.320, -0.5);
\fill (0.261, -0.5) circle (2pt);
\node[anchor=west, font=\scriptsize] at (0.56, -0.5) {1.298 (1.224, 1.377)\,*};
\node[anchor=west, font=\scriptsize] at (-0.93, -1.0) {Cell-clustered};
\draw[line width=0.8pt] (0.030, -1.0) -- (0.492, -1.0);
\fill (0.261, -1.0) circle (2pt);
\node[anchor=west, font=\scriptsize] at (0.56, -1.0) {1.298 (1.030, 1.636)\,*};
\node[anchor=west, font=\scriptsize\bfseries] at (-0.97, -1.6) {Patient-level adjustment};
\node[anchor=west, font=\scriptsize] at (-0.93, -2.1) {Ignoring clustering};
\draw[line width=0.8pt] (-0.015, -2.1) -- (0.076, -2.1);
\fill (0.031, -2.1) circle (2pt);
\node[anchor=west, font=\scriptsize] at (0.56, -2.1) {1.031 (0.985, 1.079)};
\node[anchor=west, font=\scriptsize] at (-0.93, -2.6) {Cell-clustered};
\draw[line width=0.8pt] (-0.060, -2.6) -- (0.121, -2.6);
\fill (0.031, -2.6) circle (2pt);
\node[anchor=west, font=\scriptsize] at (0.56, -2.6) {1.031 (0.942, 1.129)};
\node[anchor=west, font=\scriptsize\bfseries] at (-0.97, -3.2) {Adding marker-pattern strata};
\node[anchor=west, font=\scriptsize] at (-0.93, -3.7) {Ignoring clustering};
\draw[line width=0.8pt] (0.028, -3.7) -- (0.115, -3.7);
\fill (0.072, -3.7) circle (2pt);
\node[anchor=west, font=\scriptsize] at (0.56, -3.7) {1.074 (1.029, 1.122)\,*};
\node[anchor=west, font=\scriptsize] at (-0.93, -4.2) {Cell-clustered};
\draw[line width=0.8pt] (-0.008, -4.2) -- (0.151, -4.2);
\fill (0.072, -4.2) circle (2pt);
\node[anchor=west, font=\scriptsize] at (0.56, -4.2) {1.074 (0.992, 1.164)};
\node[anchor=west, font=\scriptsize] at (0.46, -4.68) {* 95\% interval excludes 1};
\draw (-0.10, -4.8) -- (0.52, -4.8);
\foreach \x/\lab in {-0.1054/0.9, 0/1.0, 0.1823/1.2, 0.3365/1.4, 0.4700/1.6} {
  \draw (\x, -4.8) -- (\x, -4.87);
  \node[anchor=north, font=\scriptsize] at (\x, -4.9) {\lab};
}
\node[anchor=north, font=\scriptsize, align=center] at (0.21, -5.35)
  {Ratio of postoperative stay, non-working-day\\against working-day surgery (log scale)};
\end{tikzpicture}
\caption{Ratios of postoperative stay in the illustrative application.
Exponentiated coefficients with 95\% intervals for surgery on a non-working
day after fracture repair; 39,939 patients; axis on the log scale. Intervals for cell-clustered fits come from a wild cluster
bootstrap. The dashed line marks a ratio of one. Asterisks mark intervals
excluding one.}
\label{fig:forest}
\end{figure}

In the application cohort, 3.4\% of the 39,939 patients were exposed. Figure~\ref{fig:forest} shows
the ratios of postoperative stay. The crude ratio was 1.298 (95\% CI
1.224--1.377), and cell-clustered errors widened the interval fourfold
(1.030--1.636). Patient-level adjustment moved the ratio to 1.031
(0.942--1.129 with cell-clustered errors). Adding marker-pattern strata
fixed effects moved it to 1.074, with an interval of 1.029--1.122 when
clustering was ignored and 0.992--1.164 with cell-clustered errors. The
marker-pattern strata explained 0.137 of the exposure variance. Of 2,916
strata, 404 contained both exposure levels. They held 22,772 patients
(57.0\%), including 1,304 of the 1,366 exposed patients (95.5\%).

\label{pk:45}

\section{Discussion}

\label{pk:46}

Billing items gated by facility accreditation carried enough of the treating
hospital to build strata that stood for hospitals rather than patients.
Nesting governed variance estimation: the marker-based groupings lay on
the curve traced by splitting the true hospitals at random
(Figure~\ref{fig:nesting}), so their excess type I error came from
splitting hospitals. Resolution governed adjustment and treatment
availability: the marker pattern agreed poorly with the hospitals (adjusted
Rand index 0.108--0.345) yet removed half of the confounding, because
adjustment requires only that hospitals differing on the confounder fall in
different groups, not that each hospital stays whole.

\label{pk:47}

Clustering at a coarser level than the hospital keeps cluster-robust
inference valid~\citep{cameron2015}, and the wild cluster
bootstrap~\citep{mackinnon2017} compensates for the few, unequal clusters
that undermine normal critical values~\citep{carter2017}, here bringing the
supplied-attribute cells to the nominal rate. A proxy that merges hospitals
inherits these properties, whereas one that splits them reproduces the
inflation caused by misspecified cluster membership~\citep{desai2013}. The
random hospital groups, which carried no facility information, did better
than the latent classes and reached the nominal rate as the cells did.
Standard errors are therefore best clustered on the supplied-attribute cells
with a wild cluster bootstrap; where no attribute is supplied, the latent
classes were the best available, though still inflated.

\label{pk:48}

Existing methods for adjustment assume that each patient's hospital is
recorded~\citep{localio2001}. Where it is not, studies have set the facility
aside~\citep{takano2025,iwagami2024} or adjusted for the bed-size
field~\citep{noda2025,nakamoto2026,bito2026}, the counterpart of the
supplied-attribute cells, which removed 8\% of the confounding and left
98\% of patients in mixed groups. The marker pattern removed half of the
confounding and halved the proportion in mixed groups. This is consistent
with Table~\ref{tab:gradient}, in which the strata explained most of the
variation in characteristics determined by the hospital, such as
accreditation for generic drug use, but little of that in characteristics
determined by the patient, such as sex. For adjustment and treatment availability with
respect to facility characteristics aligned with the accreditation system,
the marker pattern is therefore worth considering.

\label{pk:49}

Whether the framework can be applied to another database depends on
three measurable properties. \emph{Specificity}, on which the framework
relies, held in both settings. \emph{Sensitivity} degraded inference
gradually, and its measured value locates a database on the simulated
curve. With a fixed marker list, it was 0.742 within fee-schedule periods
(0.649--0.824 across them) but 0.488 pooled over nine years of revisions, which supports
building the strata within each period. \emph{Separability} can be computed
from public registers (2.2 hospitals per profile at cohort scale). It bounds
adjustment and treatment availability but matters little for variance
estimation, because shared profiles merge hospitals rather than split them.
Under a thinned catalogue, the type I error rate of the marker pattern
stayed within about two points of the true hospitals down to 7.4 hospitals
per profile, rising only when the hospitals fell into fewer than 30 groups
(Additional file 1: Table S4).

\label{pk:50}

This study has several limitations. First, sensitivity in the application
database could not be measured directly for lack of a facility identifier,
although a completeness-based estimate (0.741) was close to the value in
the validation database. Second, the results for adjustment and treatment
availability rest on simulated null exposures paired with real confounders
from withheld codes; they describe the groupings' analytic properties, not
their validity for real exposures. Third, the marker-based groupings may
perform worse where more patients carry no marker than the 2.1\% in the
application cohort. Fourth, with binary exposures and rare outcomes, strata
of median size 10 may often be uninformative; in the illustrative
application, 43\% of patients were in strata with a single exposure level. Fifth, an accreditation that
a hospital never bills leaves no marker and does not lower the measured
sensitivity; such a hospital is grouped with non-accredited ones, to the
detriment of adjustment and treatment availability but not of variance
estimation.

\label{pk:51}

\section{Conclusions}

\label{pk:52}

A facility proxy has no single standard of validity: nesting governs
variance estimation, and resolution governs adjustment and treatment
availability. Where no facility identifier is supplied, the
supplied-attribute cells can carry the standard errors with a wild cluster
bootstrap, and marker-pattern strata are worth considering for facility
characteristics aligned with the accreditation system.

\label{pk:53}

\section*{List of abbreviations}
ADEMP, aims, data-generating mechanisms, estimands, methods and
performance measures; CI, confidence interval; CR1, cluster-robust
variance estimator with small-sample correction; DPC, Diagnosis Procedure
Combination; ICD-10, International Classification of Diseases, 10th
Revision; IQR, interquartile range; SE, standard error.

\label{pk:54}

\section*{Declarations}

\label{pk:55}

\subsection*{Ethics approval and consent to participate}
Both databases were purchased in anonymised form without access to
identifiable information. Research involving only information already
anonymised, with no correspondence table, falls outside the scope of the
Ethical Guidelines for Medical and Biological Research Involving Human
Subjects (Japan, 2023; Chapter 1, Section 3). This study therefore did not
require ethics committee review or informed consent. The use of the data
and the disclosure of information underwent internal review at each data
provider.

\label{pk:56}

\subsection*{Consent for publication}
Not applicable.

\label{pk:57}

\subsection*{Availability of data and materials}
The data that support the findings of this study are available from Medical
Data Vision Co., Ltd.\ and DeSC Healthcare, Inc., but restrictions apply to
the availability of these data, which were used under licence for the
current study, and so are not publicly available. Data may be obtained from
the respective vendors under licence. The accreditation registers are
publicly downloadable from the regional Bureaus of Health and
Welfare~\citep{kouseikyoku2026} (downloaded July 2026). Analysis code
is available from the corresponding author on reasonable request.

\label{pk:58}

\subsection*{Competing interests}
The authors declare that they have no competing interests.

\label{pk:59}

\subsection*{Funding}
This work was supported by JST ASPIRE, Japan [grant number JPMJAP2403]; the
Tateisi Science and Technology Foundation; and a Health and Labour Sciences
Research Grant for Research on Intractable Diseases from the Ministry of
Health, Labour and Welfare of Japan [grant number 23FC1027]. The funders had
no role in the design of the study; in the collection, analysis and
interpretation of data; in the writing of the manuscript; or in the decision
to submit the article for publication.

\label{pk:60}

\subsection*{Authors' contributions}
YN and TM contributed equally to this work. Conceptualization: YN, TM;
methodology: YN, RK, AS; software: YN, TM; validation: RK; formal analysis:
YN, TM; investigation: YN, TM; resources: WF, AS; data curation: YN, TM;
writing---original draft: YN, TM; writing---review and editing: all authors;
visualization: YN; supervision: KK, AD, WF, AS; project administration: YN,
TM; funding acquisition: WF, KK. All authors read and approved the final
manuscript.

\label{pk:61}

\subsection*{Acknowledgements}
We thank the staff of Medical Data Vision Co., Ltd.\ and DeSC Healthcare,
Inc.\ for confirming, before submission, that the content of the
manuscript could be disclosed.

\label{pk:62}

\section*{Additional files}
\textbf{Additional file 1.} File format: PDF. Title: Supplementary material.
Description: cohort-selection codes (Table S1), the marker list (Table S2),
patient flow (Figure S1), the type I error rate by intraclass correlation
(Table S3), separability of the accreditation profiles (Table S4), fixed
effects crossed with the clustering of the standard errors (Table S5) and
the six withheld codes (Table S6).

\label{pk:63}

\noindent\textbf{Additional file 2.} File format: PDF. Title: ADEMP
description of the simulation studies. Description: aims,
data-generating mechanisms, estimands, methods and performance measures of
the two simulation studies, with the numbers of replications, random
seeds, handling of undefined estimates and software, each mapped to the
section of the main text or Additional file 1 where it is reported.

\label{pk:64}
\bibliographystyle{unsrtnat}
\bibliography{refs}

\begin{thebibliography}{27}
\providecommand{\natexlab}[1]{#1}
\providecommand{\url}[1]{\texttt{#1}}
\expandafter\ifx\csname urlstyle\endcsname\relax
  \providecommand{\doi}[1]{doi: #1}\else
  \providecommand{\doi}{doi: \begingroup \urlstyle{rm}\Url}\fi

\bibitem[Iwagami et~al.(2024)Iwagami, Inokuchi, Kawakami, Yamada, Goto, Kuno,
  et~al.]{iwagami2024}
Masao Iwagami, Ryota Inokuchi, Eiryo Kawakami, Tomohide Yamada, Atsushi Goto,
  Toshiki Kuno, et~al.
\newblock {Comparison of machine-learning and logistic regression models for
  prediction of 30-day unplanned readmission in electronic health records: A
  development and validation study}.
\newblock \emph{PLOS Digit Health}, 3\penalty0 (8):\penalty0 e0000578, 2024.
\newblock \doi{10.1371/journal.pdig.0000578}.

\bibitem[Localio et~al.(2001)Localio, Berlin, Ten~Have, and
  Kimmel]{localio2001}
A.~Russell Localio, Jesse~A. Berlin, Thomas~R. Ten~Have, and Stephen~E. Kimmel.
\newblock {Adjustments for Center in Multicenter Studies: An Overview}.
\newblock \emph{Ann Intern Med}, 135\penalty0 (2):\penalty0 112--123, 2001.
\newblock \doi{10.7326/0003-4819-135-2-200107170-00012}.

\bibitem[Merlo et~al.(2006)Merlo, Chaix, Ohlsson, Beckman, Johnell, Hjerpe,
  et~al.]{merlo2006}
Juan Merlo, Basile Chaix, Henrik Ohlsson, Anders Beckman, Kristina Johnell, Per
  Hjerpe, et~al.
\newblock {A brief conceptual tutorial of multilevel analysis in social
  epidemiology: using measures of clustering in multilevel logistic regression
  to investigate contextual phenomena}.
\newblock \emph{J Epidemiol Community Health}, 60\penalty0 (4):\penalty0
  290--297, 2006.
\newblock \doi{10.1136/jech.2004.029454}.

\bibitem[Ntani et~al.(2021)Ntani, Inskip, Osmond, and Coggon]{ntani2021}
Georgia Ntani, Hazel Inskip, Clive Osmond, and David Coggon.
\newblock {Consequences of ignoring clustering in linear regression}.
\newblock \emph{BMC Med Res Methodol}, 21:\penalty0 139, 2021.
\newblock \doi{10.1186/s12874-021-01333-7}.

\bibitem[Takano et~al.(2025)Takano, Kanda, Wakimoto, Ohbe, and
  Nakamura]{takano2025}
Hayabusa Takano, Naoki Kanda, Yuji Wakimoto, Hiroyuki Ohbe, and Kensuke
  Nakamura.
\newblock {Intravenous immunoglobulin for mortality and inflammatory status in
  patients with sepsis: a retrospective database study}.
\newblock \emph{Front Immunol}, 15:\penalty0 1511481, 2025.
\newblock \doi{10.3389/fimmu.2024.1511481}.

\bibitem[Noda et~al.(2025)Noda, Takahashi, Morisaki, Nishiya, Inno, Kawase,
  et~al.]{noda2025}
Kazuki Noda, Yosuke Takahashi, Akimasa Morisaki, Kenta Nishiya, Goki Inno,
  Takumi Kawase, et~al.
\newblock {Robotic-assisted versus minimally invasive mitral valve repair: a
  claims data analysis of clinical and cost outcomes}.
\newblock \emph{J Thorac Dis}, 17\penalty0 (10):\penalty0 7740--7747, 2025.
\newblock \doi{10.21037/jtd-2025-1086}.

\bibitem[Nakamoto et~al.(2026)Nakamoto, Ohata, Yuda, Fukui, Makino, Kanamori,
  et~al.]{nakamoto2026}
Daisuke Nakamoto, Emi Ohata, Mayumi Yuda, Nobuyuki Fukui, Nao Makino, Rie
  Kanamori, et~al.
\newblock {Chronic comorbidities, concomitant medications, and treatment
  patterns of chronic hepatitis B patients receiving nucleos(t)ide analogue
  therapy: a hospital claims database study in Japan}.
\newblock \emph{BMC Infect Dis}, 26\penalty0 (1):\penalty0 254, 2026.
\newblock \doi{10.1186/s12879-026-12517-1}.

\bibitem[Bito et~al.(2026)Bito, Takahashi, Kawai, and Shintani]{bito2026}
Tsubasa Bito, Shinji Takahashi, Ryota Kawai, and Ayumi Shintani.
\newblock {Association between additional non-weekday rehabilitation and
  discharge function after hip fracture, modified by age and admission
  function: a retrospective study}.
\newblock \emph{Ann Phys Rehabil Med}, 69\penalty0 (2):\penalty0 102035, 2026.
\newblock \doi{10.1016/j.rehab.2025.102035}.

\bibitem[Hammill et~al.(2009)Hammill, Hernandez, Peterson, Fonarow, Schulman,
  and Curtis]{hammill2009}
Bradley~G. Hammill, Adrian~F. Hernandez, Eric~D. Peterson, Gregg~C. Fonarow,
  Kevin~A. Schulman, and Lesley~H. Curtis.
\newblock {Linking inpatient clinical registry data to Medicare claims data
  using indirect identifiers}.
\newblock \emph{Am Heart J}, 157\penalty0 (6):\penalty0 995--1000, 2009.
\newblock \doi{10.1016/j.ahj.2009.04.002}.

\bibitem[Desai et~al.(2013)Desai, Bryson, and Robinson]{desai2013}
Manisha Desai, Susan~W. Bryson, and Thomas Robinson.
\newblock {On the use of robust estimators for standard errors in the presence
  of clustering when clustering membership is misspecified}.
\newblock \emph{Contemp Clin Trials}, 34\penalty0 (2):\penalty0 248--256, 2013.
\newblock \doi{10.1016/j.cct.2012.11.006}.

\bibitem[{Regional Bureaus of Health and Welfare, Ministry of Health, Labour
  and Welfare, Japan}(2026)]{kouseikyoku2026}
{Regional Bureaus of Health and Welfare, Ministry of Health, Labour and
  Welfare, Japan}.
\newblock Lists of accepted notifications of facility standards, 2026.
\newblock URL \url{https://kouseikyoku.mhlw.go.jp}.
\newblock Accessed July 2026.

\bibitem[Laurent et~al.(2022)Laurent, Simeone, Kuwatsuru, Hirano, Graham,
  Wakabayashi, et~al.]{laurent2022}
Thomas Laurent, Jason Simeone, Ryohei Kuwatsuru, Takahiro Hirano, Sophie
  Graham, Ryozo Wakabayashi, et~al.
\newblock {Context and Considerations for Use of Two Japanese Real-World
  Databases in Japan: Medical Data Vision and Japanese Medical Data Center}.
\newblock \emph{Drugs Real World Outcomes}, 9\penalty0 (2):\penalty0 175--187,
  2022.
\newblock \doi{10.1007/s40801-022-00296-5}.

\bibitem[Yasunaga(2025)]{yasunaga2025desc}
Hideo Yasunaga.
\newblock {DeSC database}.
\newblock \emph{Ann Clin Epidemiol}, 7\penalty0 (2):\penalty0 46--49, 2025.
\newblock \doi{10.37737/ace.25006}.

\bibitem[Nonomiya et~al.(2025)Nonomiya, Morimatsu, Kuriu, Kabata, and
  Shintani]{nonomiya2025}
Yuta Nonomiya, Hiroshi Morimatsu, Kaori Kuriu, Daijiro Kabata, and Ayumi
  Shintani.
\newblock {A new acute pain service fee addition in Japan: a nationwide study
  based on a reimbursement claims database}.
\newblock \emph{J Anesth}, 39\penalty0 (5):\penalty0 719--730, 2025.
\newblock \doi{10.1007/s00540-025-03522-w}.

\bibitem[Matsuura et~al.(2025)Matsuura, Fukushima, and Ohfuji]{matsuura2024}
Tomoka Matsuura, Wakaba Fukushima, and Satoko Ohfuji.
\newblock Estimation of the nationwide number of patients and the epidemiology
  of idiopathic osteonecrosis of the femoral head using a commercial claims
  database [in {Japanese}], 2025.
\newblock URL \url{https://plaza.umin.ac.jp/~nanbyouekigaku/}.
\newblock Accessed 12 Aug 2026.

\bibitem[Goodman(1974)]{goodman1974}
Leo~A. Goodman.
\newblock {Exploratory latent structure analysis using both identifiable and
  unidentifiable models}.
\newblock \emph{Biometrika}, 61\penalty0 (2):\penalty0 215--231, 1974.
\newblock \doi{10.1093/biomet/61.2.215}.

\bibitem[Nylund et~al.(2007)Nylund, Asparouhov, and Muth{\'e}n]{nylund2007}
Karen~L. Nylund, Tihomir Asparouhov, and Bengt~O. Muth{\'e}n.
\newblock {Deciding on the Number of Classes in Latent Class Analysis and
  Growth Mixture Modeling: A Monte Carlo Simulation Study}.
\newblock \emph{Struct Equ Modeling}, 14\penalty0 (4):\penalty0 535--569, 2007.
\newblock \doi{10.1080/10705510701575396}.

\bibitem[Sinha et~al.(2021)Sinha, Calfee, and Delucchi]{sinha2021}
Pratik Sinha, Carolyn~S. Calfee, and Kevin~L. Delucchi.
\newblock {Practitioner's Guide to Latent Class Analysis: Methodological
  Considerations and Common Pitfalls}.
\newblock \emph{Crit Care Med}, 49\penalty0 (1):\penalty0 e63--e79, 2021.
\newblock \doi{10.1097/CCM.0000000000004710}.

\bibitem[Carter et~al.(2017)Carter, Schnepel, and Steigerwald]{carter2017}
Andrew~V. Carter, Kevin~T. Schnepel, and Douglas~G. Steigerwald.
\newblock {Asymptotic Behavior of a \textit{t}-Test Robust to Cluster
  Heterogeneity}.
\newblock \emph{Rev Econ Stat}, 99\penalty0 (4):\penalty0 698--709, 2017.
\newblock \doi{10.1162/REST_a_00639}.

\bibitem[Charlson et~al.(1987)Charlson, Pompei, Ales, and
  MacKenzie]{charlson1987}
Mary~E. Charlson, Peter Pompei, Kathy~L. Ales, and C.~Ronald MacKenzie.
\newblock {A new method of classifying prognostic comorbidity in longitudinal
  studies: Development and validation}.
\newblock \emph{J Chronic Dis}, 40\penalty0 (5):\penalty0 373--383, 1987.
\newblock \doi{10.1016/0021-9681(87)90171-8}.

\bibitem[Quan et~al.(2005)Quan, Sundararajan, Halfon, Fong, Burnand, Luthi,
  et~al.]{quan2005}
Hude Quan, Vijaya Sundararajan, Patricia Halfon, Andrew Fong, Bernard Burnand,
  Jean-Christophe Luthi, et~al.
\newblock {Coding Algorithms for Defining Comorbidities in ICD-9-CM and ICD-10
  Administrative Data}.
\newblock \emph{Med Care}, 43\penalty0 (11):\penalty0 1130--1139, 2005.
\newblock \doi{10.1097/01.mlr.0000182534.19832.83}.

\bibitem[Hubert and Arabie(1985)]{hubert1985}
Lawrence Hubert and Phipps Arabie.
\newblock {Comparing partitions}.
\newblock \emph{J Classif}, 2\penalty0 (1):\penalty0 193--218, 1985.
\newblock \doi{10.1007/BF01908075}.

\bibitem[Morris et~al.(2019)Morris, White, and Crowther]{morris2019}
Tim~P. Morris, Ian~R. White, and Michael~J. Crowther.
\newblock Using simulation studies to evaluate statistical methods.
\newblock \emph{Stat Med}, 38\penalty0 (11):\penalty0 2074--2102, 2019.
\newblock \doi{10.1002/sim.8086}.

\bibitem[Cameron and Miller(2015)]{cameron2015}
A.~Colin Cameron and Douglas~L. Miller.
\newblock {A Practitioner's Guide to Cluster-Robust Inference}.
\newblock \emph{J Hum Resour}, 50\penalty0 (2):\penalty0 317--372, 2015.
\newblock \doi{10.3368/jhr.50.2.317}.

\bibitem[MacKinnon and Webb(2017)]{mackinnon2017}
James~G. MacKinnon and Matthew~D. Webb.
\newblock {Wild Bootstrap Inference for Wildly Different Cluster Sizes}.
\newblock \emph{J Appl Econom}, 32\penalty0 (2):\penalty0 233--254, 2017.
\newblock \doi{10.1002/jae.2508}.

\bibitem[Boutera et~al.(2020)Boutera, Dybvik, Hallan, and
  Gjertsen]{boutera2020}
Andrea Boutera, Eva Dybvik, Geir Hallan, and Jan-Erik Gjertsen.
\newblock {Is there a weekend effect after hip fracture surgery? A study of
  74,410 hip fractures reported to the Norwegian Hip Fracture Register}.
\newblock \emph{Acta Orthop}, 91\penalty0 (1):\penalty0 63--68, 2020.
\newblock \doi{10.1080/17453674.2019.1683945}.

\bibitem[Mori et~al.(2025)Mori, Tarasawa, Tanaka, Mori, Fushimi, Aizawa,
  et~al.]{mori2025}
Yu~Mori, Kunio Tarasawa, Hidetatsu Tanaka, Naoko Mori, Kiyohide Fushimi,
  Toshimi Aizawa, et~al.
\newblock {Limited impact of weekend admissions on hip fracture outcomes in
  elderly patients: A study from a Japanese nationwide medical claims
  database}.
\newblock \emph{Geriatr Gerontol Int}, 25\penalty0 (1):\penalty0 75--81, 2025.
\newblock \doi{10.1111/ggi.15041}.

\end{thebibliography}

\end{document}